# Problems and Solutions of Service Architecture in Small and Medium Enterprise Communities

Agustinus Andriyanto
Deakin University, Geelong, Australia
School of Information Technology
aandriya@deakin.edu.au

Robin Doss
Deakin University, Geelong, Australia
School of Information Technology
robin.doss@deakin.edu.au

*Abstract:* Lack of resources is a challenge for small and medium enterprises (SMEs) in implementing an IT-based system to facilitate more efficient business decisions and expanding the market. A community system based on service-oriented architecture (SOA) can help SMEs alleviate this problem. This paper explores and analyses the frameworks proposed by previous studies in the context of inter-enterprise SOA for SMEs. Several problems being the background of the system implementation are identified. Afterward, the offered solutions are presented, including the system architecture, technology adoption, specific elements, and collaboration model. The study also discusses the system architecture patterns of the reviewed studies as well as the collaboration organizational structures.

*Keywords*: service-oriented architecture, SOA, microservices, small and medium business, collaboration, community.

## 1. Introduction

While the Small and Medium Enterprises (SMEs) commonly have many constraints in terms of resources and capabilities, they significantly contribute to a national economy. A report from the G20 summit in 2017 stated that in most countries, more than 50 percent of gross domestic product (GDP) and two-thirds of formal employment are generated by this sector [1]. However, the report also says that adopting new technologies is one of the SME main challenges compared to the larger firms.

Service-oriented Architecture (SOA) is a means of constructing software systems based on service component modularity. As with other technologies, the SMEs face several difficulties in implementing the architecture due to many constraints, e.g., limited budget, unskilled staff, and limited IT asset and infrastructure. Some studies offer various solutions for such problem by adopting open source, migrating to the cloud platform, applying model-driven development as well as employing agile and simplified software development methodologies. Instead of managing an SOA solution individually, some studies propose a collective community system. The collaboration among SMEs not only will alleviate the limited resource issues but will also strengthen SMEs community capabilities as a whole and allows each SMEs to undertake their best competencies for more efficient collective work. This paper presents a literature review on this topic, specifically in discussing the problems, technical solutions, and collaboration of SME communities to design and implement SOA in an inter-enterprise context. The methodology and part of the result have been presented in [2], while this study examines the findings of the 37 reviewed papers in more depth.

The four common types of SOA are service architecture, service composition architecture, service inventory architecture, and service-oriented enterprise architecture [3]. In addition to these categories, the architecture also extends beyond an organization environment, labeled as an "inter-enterprise SOA" in this study. This architectural type in the context of SMEs community is the focus as depicted in fig. 1. A review was performed on 37 papers identified to be related to the topic. This paper is organized as follows. Section 2 introduces the concepts of SOA, microservices, SME, and the SME community. The next part presents the problems faced by SMEs and their communities for implementing service-based systems. Section 4 and section 5 discuss the architectural solutions and collaboration models proposed by the reviewed studies. After offering some recommendation in section 6, the paper is concluded with a summary.

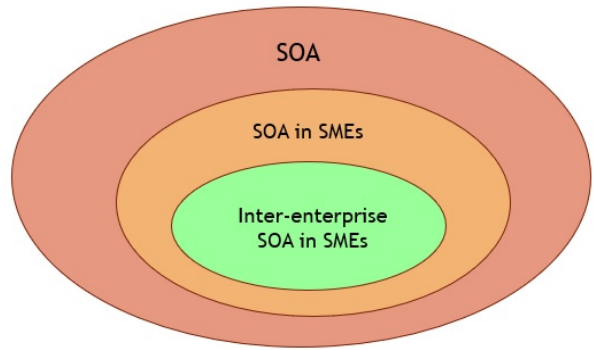

Figure 1. The focus of this literature review study

## 2. Service Architecture and SME Community

*A. SOA and Microservices*

SOA is defined as a representation of automation logic comprising of smaller, distinct units of logic which are distributed individually but collectively cover a larger piece of business automation logic [4]. When this IT architectural style was proposed several years ago, this architecture was expected to solve the integration and connectivity problem between different applications, both within an organization and externally. The service provider (SP), service registry (SR), and service consumer (SC) are the three main building blocks of SOA with their respective types of operation to interact one another: publishing for SP-SR interaction, discovery or finding for SR-SC interaction, and invocation or binding for SP-SC interaction.

SOA is viewed as a continuation of evolution in software engineering approaches. In the development paradigm, it is the latest approach after the procedural-oriented, object-oriented, and component-oriented programming. The architectural adoption means a refactoring of the system into several services that can be published by SP, maintained in repositories by the SR, as well as discovered and invoked by SC. An alternative perspective sees SOA as a newer form of software architecture which transforms the concept from monolithic, to 2-tier, and toward n-tier architecture. From the distributed programming perspective, SOA is considered as the next evolution from previous approaches, i.e., low-level socket programming, remote procedure call (RPC) programming, and distributed object computing (DOC) programming. SOA is also considered as a substantial shift in the software procurement means, i.e., from built application (the 60s-80s), to bought application (the 90s-00s), and then to composed application (the 00s-now).

The microservices architecture (MSA) has principles which are similar to SOA tenets but fix some of the problems, such as a large ESB, unwieldy configurations and scalability issues. A microservice as the basic building block of MSA is a small, independent, and single responsibility application [5]. Consisting of only hundreds of code lines, it performs only one specific task and can be deployed, tested, and scaled individually. Nevertheless, instead of contrasting MSA with SOA, the MSA can also be regarded as a specific approach of SOA;

similar to how Extreme Programming (XP) or Scrum can be viewed as specific agile software development approaches [6]. MSA is also generally regarded as a proper SOA approach compatible with recent software engineering practices and newer techniques in web application development [7].

B. *SMEs and ICT adoption*

Small and medium enterprises (SMEs) are characterized by the number of employees, total net assets, total annual sales or maximum loan sizes. The World Bank categorizes an enterprise to be an SME with under 250 employees [8]. However, different countries specify different maximum employee numbers as well as other criteria. SMEs are also distinguished with large companies by qualitative characteristics like management, personnel or finance [9] as described in table 1.

Table 1. Qualitative indicators contrasting SMEs and large companies

| Category | SMEs | Large Companies |
|---|---|---|
| Management | • Proprietor-entrepreneurship<br>• Functions linked to personalities | • Manager-entrepreneurship<br>• Division of labour by subject matters |
| Personnel | • Lack of university graduates<br>• All-round knowledge | • The dominance of university graduates<br>• Specialization |
| Organization | • Highly personalized contacts | • Highly formalized communication |
| Sales | • Competitive position not defined and uncertain | • Strong competitive position |
| Buyer's relationships | • Unstable | • Based on long-term contracts |
| Production | • Labor intensive | • Capital intensive, economies of scale |
| Research development | • Following the market, intuitive approach | • Institutionalized |
| Finance | • Role of family funds, self-financing | • Diversified ownership structure, access to anonymous capital market |

Source: UNIDO (United Nations Industrial Development Organization), as quoted in [9].

Although an ICT implementation generally leads to benefits for SMEs, some constraints are associated with the SMEs characteristics. Among others, the barriers are the lack of human capital, support by government, security, infrastructure, and financial [10]. Furthermore, strategies and tools for ICT adoption in SMEs should facilitate wider access to international markets, increase customer base, and robust information as the outcomes [10]. The next section will elaborate more the specific problems in carrying out inter-enterprise SOA in SME communities.

C. *SOA Implementation in SMEs and Inter-enterprise SOA*

When SOA was introduced, business processes in SMEs were thought not to be big enough to be implemented in SOA. SMEs are also considered to be lacking in resources to sufficiently support the whole SOA lifecycles. However, since the internet and cloud computing become more familiar while the SMEs are also required to integrate different systems either internally or externally, the service-based system becomes more prominent in today's' interconnected world.

Several studies discussed SOA implementation in SMEs. Customizing the development lifecycles is one of the ways to fit SOA with the SME situations, e.g., by simplification and reduce process steps [11] or combining strengths of BPM and SOA [12]. Meanwhile, the development and operational cost can be reduced, such as using open source sub-systems or cloud platform [13, 14]. A study by [15] presents a comparison of SOA evolution models between large enterprises and SMEs. Adopting SOA in big companies follows an '*inside-out*' transition in which IT internal development teams initially build the core services. Then, some of these services may be outsourced to external service providers, and finally, the companies can also provide services to their partners. In contrast, the realization of SOA in SMEs is modeled as an '*outside-in*' approach. SMEs pick and choose appropriate services to build solutions that would look like a mashup in the Web 2.0. The reuse is happening across whole economic environments rather than from within or across departments in a big enterprise.

All of the proposed solutions above are trying to solve the problems as an individual company. Nevertheless, a group of firms can build and manage SOA that spans among companies, i.e., the inter-enterprise SOA. This architecture is suitable to be implemented in SME communities as the development and operational cost can be more efficiently shared among the members, – a characteristic of *external economies of scale* [16]. In this case, SMEs increases their efficiency since by clustering into a bigger industry. Joining into the community will also be easier for new members because they will not be obligated to provide particular basic services or infrastructures.

## 3. Problems in SMEs and The Communities

### A. Overall Problems in The Studies

The collected papers raise some problems as the motivational background for their proposed frameworks. The number, type, and extent of the issues discussed in each study are varied. Some papers begin their discussions with problems in individual SMEs and then suggest the collaboration or inter-enterprise system alleviate the issues. Other studies start with particular matters in community scope for introducing their solutions. Therefore, the problems are partitioned into two broad groups, i.e., individual SME problems and SME community problems. Some studies cover both categories; thus, the situational overview is more comprehensively. The issues in each category then can be differentiated into non-technical and technical context, resulting in four distinct problem groups in this study. The technical context refers to any problems that purely in the developer domain.

Fig. 2 shows the top three problems in each category. Mentioned as one of the main problems in most of the groups, the issue of limited resources/infrastructure is the most reported problem for many reviewed studies. Meanwhile, heterogeneity and complexity are also regarded as serious issues in that they are addressed by two groups.

|  | Non-technical | Technical |
|---|---|---|
| Individual SME | • Limited resources;<br>• Environment;<br>• Organization; | • Agility/flexibility;<br>• Limited infrastructure;<br>• Complexity; |
| SME Community | • Collaboration;<br>• Limited resources;<br>• Heterogeneity; | • Heterogeneity;<br>• Complexity;<br>• Interoperability; |

Figure 2. Top three major problems for each category from reviewed studies

Based on the number of problem citation, the issue of limited resources including lack of funds, incapable IT staff, and the absence or ineffective IT department is the most common problem (see table 2).

Table 2. Top Problems in SMEs and SME-Communities for Implementing SOA

| No | Scope * | Context ** | Problem | Details | #Papers mention |
|---|---|---|---|---|---|
| 1. | I | NT | Limited resources | Limited resources in general, budget/fund, IT staff with their workforce and knowledge, the absence of limited IT department | 19 |
| 2. | C | T | Heterogeneity and data compatibility | Heterogeneity in general, software, existing data, business, semantic support, compatibility among data | 14 |
| 3. | I | NT | External, e.g. limited market | Environment changes and challenges, restricted market, limited partner | 11 |
| 4. | C | T | Complexity | Integration, service composition, data analysis | 9 |
| 5. | I | T | Agility/Flexibility | Adapting recent technologies, customizing the solution, installing and replacing services, extending the system, accommodating dynamic requirements, provisioning plug-and-play elements | 8 |
| 6. | I | NT | Organizational Issues | The solution on technological focus (not on business focus), differentiation for value-added business, business confidentiality, competitive quality, innovation culture, failure to engage in ICT and e-business | 6 |
| 7. | I | NT | Limited capabilities | Operational area, capacity and small-scale products, standard compliance, low technology utilization, technical support | 4 |
| 8. | I | NT | Independency and business continuity | Subordination of larger industry players, short-term contracts | 4 |
| 9. | C | NT | Collaboration (Non-Technical) | Sporadic and low culture collaboration, lack of information instruction, production disorder and intermittent production, virtual team limitations, leadership | 4 |
| 10. | C | T | Collaboration (Technical) | Low synchronization and visibility, dynamic collaboration support, small-scale and horizontal collaboration support, balancing control and autonomy | 4 |
| 11. | C | T | Distributed environment | General problems in the distributed environment, fragmented ICTs | 4 |

* Scope = I (Individual), C (Community)
**Context = T (Technical), NT (Non-Technical)

### B. Non-Technical Problems in Individual SME

The problems faced by SMEs not related to any technologies or methods are categorized under this group. Among all other issues, the limited resource is the most prominent problem (see fig. 3). Therefore, it is the strongest rationale for SMEs to form the inter-enterprise SOA systems. Most of the papers identify budget, IT staff, and the IT department as the main elements. The budget issue includes limited fund to purchase infrastructure, operate the system, or adopted new technologies [17], [18], [19], [20], [21], [22], [23], [24].

The external factor, in the second rank, mostly describes SMEs' challenges to survive and adapt in today's dynamic world driven by ICT that rapidly changes many aspects in society. They must strive to stay competitive, satisfy various requirements of customers rapidly, survive in a dynamically changing environment, and keep pace with the global trend [25],

[17]. It is not always easy to penetrate new markets while facing some hindrances, such as the lack of sufficient information and market signals [26] [27]. Furthermore, SMEs usually have limited partner networks outside their regions or in other countries [28]. Several papers also raise independency and business continuity issues. Although SMEs are self-governing entities, they are highly dependent on larger industry players and regarded as only "piece-workers" to do specific tasks [29], [30]. Since non-technical issues in individual SME cover general situation in SMEs' circumstances, they might be discussed either in implementing SOA or in carrying out general ICT-based system.

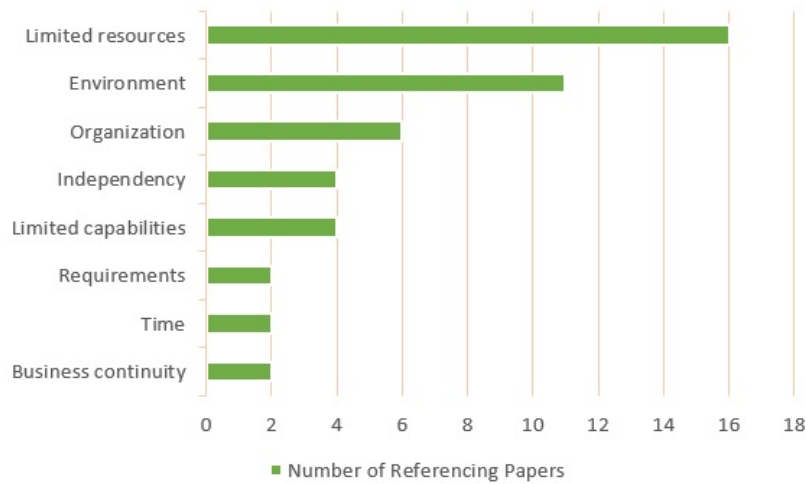

Figure 3. Non-technical problems in individual SME mentioned by the studies

## C. Technical Problems in Individual SMEs

The reviewed papers do not discuss technical issues in individual SME as much as the non-technical one. Agility or flexibility is the most discussed issue in this group which covers several cases, e.g., customization, technology adaptation, and plug-and-play elements. Agility refers to sense and response capability in adapting and performing well against rapidly changing environments [31]. Hence, SMEs need to adjust to the constant evolution and dynamics of technologies as well as regulations and requirements.

## D. Non-Technical Problems in SME Communities

As in technical problems of individual SME, there are not many non-technical problems addressed in the SME community context. The most discussed topic is about collaboration problems among the community members. Collaboration is defined as any action or work involving two or more entities take part of agreed aims by sharing their distinct resources and knowing how to achieve those aims [32]. In a situation where small and medium businesses have overwhelming difficulties in maintaining their competitiveness, the *partnership* is a good option to extend their sustainability. Nevertheless, the collaboration culture is low in that firms work on a team with the same partners and perform the actions sporadically [25]. Some industry clusters composed of small and medium-sized firms are lacking in information instruction that causes confusion where products are processed sporadically and managed disorderly [27].

The SMEs communities as groups also face limited resource issues like in an individual company. The small parties have little money to invest in ICT infrastructure and services since they are low-cost-based clusters where many of them are labor intensive and of small-scale

[33], [19]. When implementing inter-organizational business process engineering for system integration and networking, they lack a considerable budget means that low-cost technological platforms are preferred [34], [35], [36].

*E. Technical Problems in SME Communities*

To compete with large firms, SMEs organize alliances consisting of several actors, but they experience enormous diversity over their running services and infrastructures. Another problem happens for new member enterprises joining an established community while they must align their existing infrastructure, business processes, document formats, and document content. Heterogeneity, along with complexity and interoperability are the three top technical problems in SME communities (see fig. 4). Interoperability includes data compatibility and semantic support for communication and service composition. As the main SOA goals are to solve integration and interoperability issues among distributed systems, the SOA-based frameworks offered by the reviewed papers attempt to address them.

Other issues are discussed to some extent like access, distributed environment, and standard, while some issues such as security, automatic composition, and monitoring do not get much attention from the studies. Unlike the first problem group, fig. 4 shows that technical problems in SME communities are more diverse and more distributed.

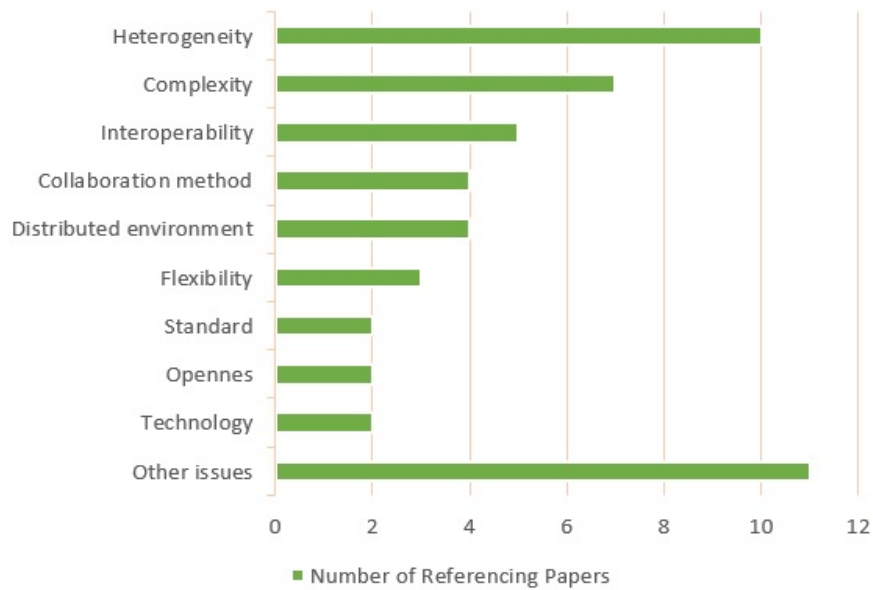

Figure 4. Technical problems in SME communities mentioned by the studies

## 4. Proposed Solutions

This paper maps the solutions and the issues based on the top 11 problems (table 2) as shown in table 3. One solution may also address several issues at once. The general solutions such as the community-based system, standards enforcement, or mediator are inherent in many proposed frameworks. Therefore, table 3 does not mention which frameworks/papers suggest which general solutions and provides only the references for the specific solutions. Apart from the proposals classified as general and specific solutions, this study also groups the proposals into four subtopics, namely architectural solutions, technology adoption, specific elements, and collaboration strategies and models.

Table 3. The solutions for problems depicted in table 2

| No  | Problem | General Solutions | Specific Solutions |
|-----|---------|-------------------|--------------------|
| 1.  | Limited resources | Community-based system, | cloud [17], [36], [37], [38], [39], resource sharing/ warehouse [40], [41], [42], [36], [37], simple peer [43], mobile technology [44], [45]. |
| 2.  | Heterogeneity and data compatibility | Mediator/ESB, standard, web service wrapper, business integration | ontology [18], [43], [22], [46], agent [44], [28], [21], |
| 3.  | External, e.g. limited market | Community-based system, business integration | Knowledge centre [47], |
| 4.  | Complexity | Suitable standard, layered system, web service wrapper, service bus | MDA [43], [38], [20], cloud [17], [36], [37], [38], [39], ontology [18], [43], [22], [46] |
| 5.  | Agility/Flexibility | Layered system, web service wrapper, service registry | MDA [43], [38], [20], EDA [23], [18], collaboration layers [25], [48] |
| 6.  | Organizational Issues, e.g. innovation culture, business differentiation | Community-based system | Socio-technical requirement formulation [29], [30], |
| 7.  | Limited capabilities | Community-based system, | resource sharing/ warehouse [40], [41], [42], [36], [37] |
| 8.  | Independency and business continuity | Community-based system | Leader/extended enterprise [34], [49] |
| 9.  | Collaboration (Non-Technical) | Community-based system, integrated workflow and workflow engine | Leader/extended enterprise [34], [49] |
| 10. | Collaboration (Technical) | Community-based system, mediator/ESB, service registry, integrated workflow and workflow engine | EDA [23], [18], collaboration layers [25], [48] |
| 11. | Distributed environment | Community-based system, mediator/ESB, service registry | EDA [23], [18] |

As the central topic of this study, the community-based system is proposed by all of the reviewed papers. Therefore, this study does not categorize the solution into one of the subtopics. Some proposals can be general approaches that focus on particular aspects like integration and interoperability [24, 40, 50] or cloud platform [17, 37]. Meanwhile, other frameworks perform specific functions, such as knowledge management [43], data warehouse [41], and distributed ERP [42, 51, 52]. Some also suggest architectures that extend the context beyond SMEs community such as e-Knowledge centre for strategic planning [47] and urban public service system [26].

### A. Architectural Solutions

#### 1) Layered System

Service layers in SOA provide many advantages, e.g., system abstraction, independency between business logic and technology-specific application logic, organization agility, and component reusability. Three layers of abstraction defined by [4] are the application service layer, the business service layer, and the orchestration service layer. Some of the reviewed studies provide various service layers regarding the number and types. Nevertheless, they have a similar pattern that covers business context on the highest level to physical/infrastructure concern on the lowest level [17, 19, 21, 39, 44]. While most of the layers are positioned horizontally on top of the other and related to their functional natures, in [44] there are three extra layers added vertically for cross-cutting concerns.

*2) Resource Sharing/Warehouse*

A group of organizations can share its distributed IT resources to construct the system quickly and economically. Extending SOA model, [40] presents a design for inter-organizational IT resources sharing for supply chain management system. The proposed implementation of the framework may be carried out by leading vendors of industry or by application service providers (ASP). Another solution, i.e., a resource warehouse proposed by [41] provides a suitable data warehousing infrastructure to SMEs which have an enormous volume of data.

The resource sharing system also includes distributed ERP solutions provisioned by multiple parties, e.g., an ERP mall [42]. The system composes many components that are distributed across different providers. SMEs as consumers select and use the services, and then pay only the used parts. Meanwhile, the roles of a mediator are in implementing the infrastructure, integrating services, providing support for the development environment and evaluation as well as consultation and training. It is also responsible for making offered services available to users, ensuring SLA compliance, and determining penalties for service providers. Another ERP system in food industry networked SMEs uses a controller to encapsulate distributed modules into cloud resources [39] and orchestrate them through a centralized SOA. The domain-specific controller selects appropriate ERP systems using a multi-criteria optimization method. Afterward, the networked SMEs can collaborate within a value chain with better resources utilization in accordance with the customer requirements. In the cloud platform, [36] and [37] present centralized public SaaS models from different software vendors to provide office automation, ERP, e-mail, supply chain planning, etc. for small and medium companies.

*3) Simple Peer*

In a P2P framework, each SME node has to provide its own infrastructure. A knowledge management system, as suggested by [43], presents a collaboration within a digital business ecosystem environment (DBE) for discovering, retrieving, and transforming particular data. Since the environment is P2P, SME nodes in the community must deploy the knowledge base (KB) infrastructure for enabling the three core services: KB service, semantic registry service, and recommender service. However, this unstructured P2P paradigm still allows simple peers that do not wish to share resources and install the infrastructure. They are still able to use services offered by other knowledge host members as a part of the DBE.

*4) Knowledge centre*

Designed as an internet-based community system, a knowledge center help people, enterprises, and government to share and consume the information. The knowledge center in [47] is a combination of decision support system and e-knowledge services for strategic planning to promote the cooperation between local government and SME applications. It provides a strategic tool for local governments regarding territory growth while also provides marketing policy recommendations for SMEs. The integration of enormous data from the web, local government, and SMEs is performed by crawling and web mining techniques.

**B. Technology Adoption**

*1) Cloud*

Cloud computing gives many advantages to SMEs and their communities, e.g., cost savings, flexibility, security, mobility, etc. Virtualization is an essential enabler of most cloud platforms with its ability to provide an abstraction of infrastructure, operating system, or application instances. As a solution for a heterogeneously distributed infrastructure, [17] suggests a cloud system based on SOA and VE principles. The proposed platform is named as Open Nebula, an open source virtual infrastructure engine that decouples the server not only from the physical infrastructure but also from the physical location. The cloud platform transforms software providers from portal-based e-services into SaaS applications as proposed by [36].

A public SaaS platform from multiple software providers based on cloud service bus is also proposed by [37]. The bus extends the enterprise service bus (ESB) to support interoperability among integrated software. Another SaaS is SOA4BIM framework [38] where a building information model (BIM) is dynamically refined and enriched by different parties. Meanwhile, using cloud architecture for distributed ERP implementation becomes more common like in [39] that presents optimization and control in networked food industry SMEs for operating a collaborative ERP system.

*2) Mobile Technology*

Mobile technology enables flexible workplaces and lower infrastructure cost that result in higher efficiency and productivity. Several papers propose mobile technology, such as in [44], that includes an intelligent agent as a mediator in a peer-to-peer architecture. Mobile proxies could also act as agent servers for mobile terminals. As mobile devices are resource constrained, the web services use REST rather than SOAP/WSDL. Another framework proposition deploys a service directory on each mobile device in a P2P environment [45].

*3) Web Services and Standards*

Although services in SOA can be implemented by different platforms, mostly they are realized by web services. Therefore, many SOA applications refer to the web service standards with two mainstream technologies, i.e., SOAP and REST. SOAP is based on XML (eXtensible Markup Language) for communication between participants. This XML-based protocol consists of three basic standards: SOAP (Simple Object Access Protocol) for exchanging information between distributed applications, WSDL (Web Services Description Language) for describing web services, and UDDI (Universal Description, Discovery, and Integration) for publishing and finding web services.

The other protocol, i.e., REST, is a newer architecture that simplifies the request and message processing method and uses only HTTP as the application-layer protocol. A study by [53] shows that the RESTful style services offer better performance than the SOAP-based services. However, most of the reviewed studies still make use of SOAP to interact between parties.

For exchange of business data, there are some standards, such as EDI (electronic data interchange), ebXML, or RosettaNet. Enforcing a same guideline in a community can simplify the solution and alleviate heterogeneity problems. Nevertheless, it may cost additional effort for the community members to adopt the standards set.

*4) Model-driven Architecture (MDA)*

MDA is a standard of model-driven development (MDD) methodology defined by the Object Management Group (OMG). In MDD, software developers construct a system model and then transform it into real solution components. The OMG defines MDA as a method of describing the IT system in which system functionality specification is separated from its implementation on a particular technology platform [54]. Three models are defined by the MDA for abstracting a system into different viewpoint levels, i.e., Computation Independent Model (CIM), Platform Independent Model (PIM), and Platform Specific Model (PSM).

The knowledge management platform proposed by [43] applies MDA to build the kernel of supporting semantic services in the digital business ecosystem. Meanwhile, [38] utilizes MDA to develop the Building Information Model (BIM) cloud services accessed by applications or web end users. The owner team and an architect create the initial model and program to the cloud. Evoking a cloud service by agents (client, architect, specialist designer, etc.) will transform the PIM, such as engineering analyzing, duct routing, room requirements to the specific PSM, such as web services. The agents then refine the design with their internal applications and transform it into the enriched PIM model. Another use of MDA by [20] develops business process orchestration on ESB from the Business Process Modelling Notation (BPMN) model.

*5) Event-driven Architecture (EDA)*

Although not mentioned as much as MDA, some of the studies apply EDA for their proposals. EDA is an approach in which applications and systems are designed and implemented to enable event delivery between decoupled software components and services [55]. Instead of callers must explicitly request information as though in a request/reply system, this approach allows the system to dynamically counter as events take place. One of the EDA adoptions is the event-bus [23] that comprises several sub-buses where connected stakeholders can create, disseminate, and respond to events. Another framework combines SOA, EDA, and Semantic Web Architecture (SWA) to support asynchronous communication with the objective for sharing data and business interoperability among members [18].

### C. Specific Elements

*1) Mediator/hub: ESB and Portal*

In an SOA-based system, the enterprise service bus (ESB) is a central element that acts as a mediator, hub or portal for the participating members. It performs many functionalities: communication infrastructure, reliable messaging, controlling network traffic, data format transformation, message enhancement, service management, etc. The key capabilities of an ESB are summarized in [56]. Mainly, it is used in the context of application integration, ESB replaces the previous technical solutions, i.e. ETL (Extract Transform Load) and EAI (Enterprise Application Integration).

Promoting SOA for business integration involves this central element like a brokerage model for B2B cooperation between SMEs in consortium [50], information service platform [19], external integration using a visual tool [57], and business interaction between buyers and suppliers in group-buying SMEs [33]. The service bus may be implemented on the cloud for public SaaS [36, 37]. Also, the mediator may administer mobile nodes in P2P architecture that is represented as an agent [44]. Meanwhile, the SME business integration can encompass different industries in several countries or vast areas, such as e-commerce in [24] and extended logistic enterprise for distributed supply chain in [49]. The ESB can be customized to support a decentralized system by deploying it on each SME [58]. Incorporating event-driven architecture (EDA), another design of customized ESB [23] enables each stakeholder to create and disseminate an event or make a respond.

*2) Web Service Wrapper*

A web service wrapper enables legacy systems to engage in an SOA-based application with less or no modification. Therefore, a participant has more freedom to maintain services with the existing programming languages and standards. This condition leads to ease of integration and coordination, such as in [19] where web service wrappers expose the legacy systems and register them to UDDI registration center, while new applications are built based on web service technology.

*3) Ontology*

Ontology helps in solving incompatibilities among data exchanged between SMEs. It also addresses interoperability problems for integrating data and business processes in service networks. Using an ontology to represent services engaged in business processes, the framework in [46] can manage the creation, operation, and dissolution of virtual enterprises or organizations. The integration is performed through a matching process of the service (including its input and output parameters) invoked by an SME with a service implemented by another company.

Some other ontology-based systems are suggested by the reviewed studies. One of them, a knowledge management system in the P2P environment [43] relies on the knowledge base (KB) as its core infrastructural components to provide a knowledge representation. Communicating enterprises can understand each other by interoperable metadata based on *meta-object facility* (MOF) architecture from the *object management group* (OMG). Meanwhile, the ontology mechanism by [22] involves two entities, i.e., global common ontology which acts as the primary reference for information exchange between applications, and local ontology resides in each stakeholder site. An ontology translation mechanism is performed if the local ontologies of two interacting parties are different. Another ontology framework proposed by [18] defines an interoperability ontology for business concept model that shares concepts for value exchange between enterprises, such as business activity, resource, and actor. A specific syntax then actualizes the ontology, like EDI (Electronic Data Interchange) or XML format to be used in web services or EDI messages.

*4) Agent*

Having more autonomy and flexible behavior than general programs, an agent is used as parts of the system to automate processes and interactions. The agent applications in [28] and [21] promotes the virtual-organization creation by automating the exchange information, discovery and negotiation processes with the collaborative partners. Each company has one or more agents that communicate with other agents in other companies and performs negotiation based on user-defined business rules. For the requirement of managing SME data, the solution vendor hosts the central portal and approves the participants that can involve in several value chains. Another architecture in [44] uses an agent for mediating the interacting parties in P2P mobile network where each service client also acts as a service provider. In this case, the agent server is also a mobile proxy that manages a group of mobile terminals. Connected to the fixed infrastructure, which acts as a node in the P2P network.

*5) Service Registry*

The main role of the service registry is to maintain a list of services published by service providers and gives this information to service consumers who perform discovery or searching operations. A P2P framework for mobile environment [45] uses XMPP (eXtensbile Messaging and Presence Protocol) to maintain a service directory in each mobile device. In that model, the device plays all the roles in SOA, i.e., as a service provider, service registry, and service consumer or requester.

### D. Collaboration Strategies and Models

*1) Business Integration*

Integrating business processes from different SMEs will result in more efficient workflow, more scalable production, and broader market. Through dynamic virtual organization (VO), an integrated workflow involving many participating SMEs is established and then executed [48]. The integration can be B2B, such as in rail manufacturing companies [50], an extended logistic enterprise [34], or virtual hotel organization [59]. In [34], a central server stores process definitions/models in a business-process repository and orchestrate activities execution. Meanwhile, the clients perform their processes in accordance with their roles in the supply chain.

*2) Socio-technical Requirement Formulation*

Specifying requirements does not include technical matters only but also business issues, e.g., innovation culture, market opportunity, and customer satisfaction. A socio-technical approach in [29, 30] identifies three categories for the SME alliance requirements, i.e., functionality to support alliance business management, functionality to support alliance operation management, and functionality to support alliance capability, learning, and innovation. Each category is decomposed into some sub-category, e.g., for the second group, one of the sub-categories is an alliance formation and membership management, which then consists of several functionalities, such as SME member registration, membership validation, and profile definition.

*3) Integrated workflow and workflow engine*

A workflow engine orchestrates services by assigning responsibilities, specifying an order for each of the services, and execute them so that it accomplishes the set goals. The application for group-buying SMEs [33] implements a workflow engine to automate business interaction in a virtual organization between buyers and suppliers. Managing workflow is also defined in supply chain collaboration among SMEs and third party (3PL) logistic firms [27]. In this system, various SME resources are united with 3PL firms as the medium. The knowledge database plays a vital role by storing the methods, workflow docking models, and synergy operations knowledge. Likewise, there are other studies that propose collaborative workflows among SMEs with the workflow engine, e.g., federated message-based architecture for eBusiness interoperability [24] and INPREX [34].

*4) Collaboration Layers*

Breaking down a collaboration model into some layers enables a more agile partnership among enterprises. An alliance model among independent software service providers (ISP) in [25] defines four layers of the collaboration scenario, i.e.:

- VBE layer represents the ISPs long-term alliance and the basic partnership environment;
- Service cloud layer constitutes all ISPs services as the basis for creating composite solutions;
- VO (virtual organization) layer reflects possible software services solutions composed of existing services in the previous layer;
- Auxiliary partnership layer points out cooperation with external partners to support the service solution business.

The first and the third layers of this framework resemble the static base VO and the dynamic VO defined by [48].

*5) Leader/Extended Enterprise*

Although giving many advantages, a collaboration among SMEs sometimes needs a strong leader or a large company to ensure credibility, business continuity, and proper infrastructure. Therefore, one dominant company and several other smaller firms can establish an alliance named extended enterprise. A collaboration model in [49] involves a logistic enterprise that integrates some consortia partners. The dominant leader can also be a medium-size firm. Another collaboration model in business processes aggregate by [34] places SMEs as first-tier suppliers for a number of automobile manufacturers and they are responsible for their own logistics and $2^{nd}$-tiers suppliers.

*E. Other Aspects of The Frameworks*

*1) Structural architecture form*

This study adopts the types of organizational structure commonly applied in system or enterprise architecture as defined in [60] and [61], i.e., centralized, federated/hybrid, and decentralized or peer-to-peer (P2P). While in a centralized system the control is given only to one entity, in a decentralized system the control is shared among the participants. The federated system combines the characteristic of both architectures. Most of the frameworks present centralized designs, and only a few of them suggest federated or decentralized solutions.

*Centralized Architecture*

Entities in a centralized architecture are interconnected through a mediator or a central element like interoperability bus and centralized portal [19, 24, 33] or eHubs [36, 49, 57]. In a collaborative ERP system through centralized SOA, selecting the appropriate distributed modules for certain requirements may involve a specific controller that performs a multi-criteria optimization method [39]. Meanwhile, in the case of e-knowledge center for territory growth [47], the central entity integrates enormous data from the web, local government, and SMEs by crawling and web mining techniques. The central component that performs many tasks can be provided and managed by a consortium of SMEs, large firms, or government agencies. A big operator-integrator in urban public service system [26] administers the inter-community system and its backbone infrastructure.

*Federated or Hybrid Architecture*

Control in a federated system is shared between the central entity and the participants or the subcentral elements. A hybrid P2P architecture proposed by [44] employs an intelligent agent for mediation among mobile nodes in P2P architecture. Each client machine (i.e., mobile terminal) can act either as a service provider or service consumer. However, it is not pure P2P as there is a control node that serves as a proxy and an agent connected to the fixed infrastructure. In a federated ERP system [42], the central element integrates many components from many resource providers and then makes the resources available to SME consumers. Another federated architecture is a public SaaS by [37] whose service bus model depends on workload requirements. As its workload increases, the service bus model expands from one ESB server to a cluster consisting of several ESB server.

*Decentralized or Peer-to-Peer (P2P) Architecture*

The benefits of using P2P architecture is the absence of a single point of failure, as the interactions do not involve a central node. A P2P architecture with private cloud platform enables an SME community to form virtual enterprises (VEs) with all its business process services, IT services, technologies, and their providers [17]. In a digital business ecosystem environment (DBE), SME may collaborate to develop an infrastructure for knowledge-based business communities, for instance, a P2P knowledge management framework suggested by [43].

P2P architecture is also proposed in [51] and [52]. The architecture offers a distributed ERP system in which participants provide specific goods and services to SMEs. A company only needs a server with preinstalled database and basic functionalities to connect to the internet and interact with other peers for fulfilling its requirements.

2) *Collaboration Models*

All the proposals from reviewed studies are intended to allow enterprises to collaborate more efficiently, which in turn increases competitiveness in the global landscape. Despite many manifestations of collaborative networks, such as those defined by [62], this paper classifies the collaboration model among SMEs according to their basic structural form that is adapted from [63] as described below.

*P2P Virtual Enterprise and Virtual Breeding Environment*

There is no central actor in the P2P collaboration. Common virtual enterprise (VE)/virtual organization (VO) and VO breeding environment (VBE) are included in this category. In a virtual enterprise (VE), the members cooperate temporarily in computer networks with short-term goals, while in a virtual breeding environment (VBE), the members maintain a static long-term alliance and agreement with interoperable infrastructure. Most of the reviewed studies have their collaboration models under this type like in the architecture of e-commerce [24] or the application for group-buying portal [33].

*Extended Enterprise*

As previously discussed, an extended enterprise entails a central actor for the collaboration. The principal extends its business boundaries to the subordinates. While the major company can concentrate on its core competencies, the subordinates get benefits from their products or services supplied to the dominant firm.

*Supply/Value Chain*

A supply/value chain community manages the activities related to the flow of goods or services from the initial supplier to the final customer. In this model, the members interact with their immediate neighbors in the business chain to make the whole supply chain more efficient and competitive. Some frameworks within this model are integration frameworks based on a milestone model [35] and a typical supply chain collaboration between SMEs and third party (3PL) logistic firms [27].

## 5. Discussion

The reviewed studies offer frameworks in an attempt to address the problems related to service-based systems implementation either, in a community or individual context of small and medium enterprises. The lack of sufficient resources is the problem experienced by most SMEs and the communities. Therefore, this issue is the focus of most of the papers for proposing their solution frameworks. Although the reviewed studies recommend SOA-based systems in SME communities, the identified problems also cover non-technical and individual SME issues.

Different designs are proposed for a specific context. Most of the reviewed studies raise its concerns on the community-based architectural solutions along with their constituent elements. Other papers focus only on parts of the architecture, such as ESB or ontology. Nevertheless, there are many diversities in the architecture breakdown, from a simple general architecture introduction to an elaboration of detailed constructs. In addition, only a few of them present how the systems are partitioned onto some layers. It is also better if the proposed frameworks can refer to and consider several related architecture standards, e.g., SOA reference architecture from Open Group [64], reference architecture foundation for SOA from OASIS [65], and reference architecture for SOA from ISO/IEC [66].

Some technologies like MDA or EDA are employed in the proposed frameworks. While the mobile and cloud platforms have capability to reduce infrastructure cost significantly, most of the proposed solutions have not considered them. It is due to the immaturity of the technologies in the time of papers publication. Other recent trends in service delivery platforms, such as Internet of Things and edge computing have not been elaborated as well. In the aspect of web service technologies, the proposed solutions mainly implement SOAP for web services, while in reality, REST outperforms SOAP in terms of performance and simplicity.

Due to the natural characteristics of SOA and SMEs, the most recommended organizational structure of the systems by the studies is the centralized architecture. ESB (enterprise service bus) acts as a central component mediator in the typical SOA model, connecting provisioned services and their clients. Various tasks performed by the ESB includes message transformation, routing, security, monitoring, service orchestration, service management, etc.

Therefore, each SME connected to the community system do not have to build a complex application as the ESB can perform many common functionalities. By moving some complexities to the central community element, i.e., ESB. SMEs will find this integrated environment quite ideal and save their spending on resources. A centralized system is also easier to manage. On the other hand, the risk is higher as a failure of one central component can bring the entire system down. Furthermore, low scalability and low agility are the typical disadvantages of centralised architecture.

A decentralized system with more stable, scalable, and agile characteristics may solve the issues, but it also comes with some consequences as the management and monitoring are more complicated. In addition to the high complexity inherent in the endpoints (in this case SMEs), a decentralized system required more efforts to establish methods and elements standards. A federated architecture can balance the benefits and disadvantages of these two constructs. Nevertheless, integrating distributed systems has to deal with extra development efforts and complexities in considering the composed elements, either in a centralized, federated, or a decentralized system.

As the glue of the community systems, collaboration among SME members is modeled by some patterns following three basic forms: P2P virtual enterprise/virtual breeding environment, extended enterprise, and supply/value chain. What collaboration pattern must be selected depends on the business nature of the alliance itself. Nevertheless, not many papers discuss the topics about membership types in an association or collaborating with some external organizations when some capabilities are not available inside the community.

Most of the reviewed studies assume that SMEs to some extent already operate IT-based system, employ staff with good IT knowledge and skills, and good internet connection and bandwidth. However, it may not be the case for the condition in underdeveloped and developing countries. Some papers suggest the decentralized model which is coherence with the typical nature of microservices architecture (MSA). Nevertheless, these studies have not considered this architecture. Prospective studies can adopt some tenets from centralized SOA and decentralized MSA to get optimum benefits from these two concepts. In that way, centralized interoperability may be combined with a decentralized implementation for balancing the goals of simplicity at endpoints and agility in the whole system.

## 6. Conclusion

When developing an inter-enterprise system based on service-oriented architecture, SME communities face many challenges mainly in limited resources, interoperability, and environment/market changes. The reviewed studies propose some frameworks that include architectural solutions, technology adoption, specific components, or collaboration models. The choice of organizational structure of the architecture, whether it is centralized, federated, and decentralized has its own costs and benefits. Meanwhile, the basic structure of collaboration models will follow the business nature of the alliance. Future work for this research will design a service architecture that addresses several issues not yet fully elaborated, such as the interaction to external organizations outside the community, conforming with SOA reference architecture standards, and the adoption of microservices architecture.

## 7. Acknowledgment